\newcommand{\CLASSINPUTtoptextmargin}{0.65in}
\documentclass[lettersize,journal]{IEEEtran}
\usepackage{array}
\usepackage[caption=false,font=footnotesize,labelfont=rm,textfont=rm,subrefformat=parens,labelformat=parens]{subfig}
\usepackage{textcomp}
\usepackage{stfloats}
\usepackage{url}
\usepackage{verbatim}
\usepackage{graphicx}
\usepackage{cite}
\usepackage{amsmath,amssymb,amsfonts}
\usepackage{xcolor}
\usepackage{CJK}
\usepackage{svg}
\usepackage{colortbl}
\usepackage[colorlinks=true, linkcolor=blue, citecolor=blue,urlcolor=black]{hyperref}
\usepackage{booktabs}
\usepackage{steinmetz}
\usepackage{soul}
\usepackage[linesnumbered,ruled]{algorithm2e}
\usepackage{bm}
\usepackage{makecell}
\usepackage{multirow}
\usepackage{newtxtext,newtxmath}
\usepackage{tabularx}

\begin{document}
\bstctlcite{IEEEexample:BSTcontrol}

\title{\vspace{-20pt}
Joint Visible Light and RF Backscatter Communications for Ambient IoT Networks: Fundamentals, Applications, and Opportunities}

\author{Boxuan Xie, Yifan Zhang, Kalle Koskinen, Alexis A. Dowhuszko, Jiacheng Wang, Ruichen Zhang, \\Zehui Xiong, Zhu Han, and Riku Jäntti
\vspace{-30pt}
}


\maketitle

\begin{abstract}
The rapid growth of the Internet of Things (IoT) devices in the sixth generation (6G) wireless networks raises significant sustainability and scalability challenges due to energy consumption, deployment complexity, and environmental impact. Ambient IoT (A-IoT), leveraging ambient energy harvesting (EH) for batteryless device operation, has emerged as a promising solution to address these challenges. 
Among various EH and communication techniques, visible light communication (VLC) integrated with ambient backscatter communication (AmBC) offers remarkable advantages, including energy neutrality, high reliability, and enhanced security. 
In this article, we propose a joint VLC-AmBC architecture, emphasizing fundamental concepts, system designs, and practical implementations. 
We explore potential applications in environmental monitoring, healthcare, smart logistics, and secure communications. 
We present proof-of-concept demonstrations for three distinct types of ambient backscatter devices (AmBDs): EH-Only, VLC-Relay, and VLC-Control. Experimental results demonstrate the feasibility of implementing joint VLC-AmBC systems, highlighting their practical viability across various deployment scenarios. Finally, we outline future research directions, including integrated sensing and communication, as well as optimized energy-efficient deployment.
Open issues, such as large-scale deployment challenges, are also discussed, thereby providing a clear roadmap for future developments in joint VLC-AmBC-enabled A-IoT ecosystems.
\end{abstract}

\begin{IEEEkeywords}
Ambient IoT, backscatter communication, optical wireless, 3GPP, tag, energy harvesting, zero energy device.
\vspace{-15pt}
\end{IEEEkeywords}

\section{Introduction}
The era of the sixth generation (6G) wireless networks is driving a rapid growth in the use of the Internet of Things (IoT) for smart environments, healthcare, intelligent transportation, industrial automation, and more. 
However, the substantial increase in IoT devices necessitates significant energy resources to maintain their continuous operations~\cite{lopez2025zed}. 
Conventional battery-based or wired solutions for powering IoT devices are regarded as inefficient, especially considering their deployment complexity, frequent maintenance requirements, and environmental impacts from discarded batteries. 
Consequently, energy harvesting (EH), giving rise to the concept of Ambient IoT (A-IoT), has emerged as a promising batteryless IoT solution endorsed by the 3rd Generation Partnership Project (3GPP)~\cite{3gpp38848}. A-IoT envisions IoT devices operating sustainably through harvesting energy from the ambient environment, thereby achieving energy-neutral operations~\cite{kimionis2024aiot}.

Backscatter communication (BC) has emerged as a pillar technology enabling ultra-low-power wireless connectivity for A-IoT~\cite{jantti2025aiot}. In a BC system, a battery-free backscatter device~(BD) modulates and reflects incident electromagnetic waves for communication, rather than generating its own RF carrier. 
Nevertheless, conventional BC, such as passive radio-frequency identifier~(RFID) systems, suffer from limited communication ranges and necessitate dedicated readers for BD activation and message reception~\cite{jiang2023survey}. 
Such limitations constrain their application scenarios.

Recently, ambient backscatter communication (AmBC) has emerged as a promising enhancement to conventional BC~\cite{cui2025magazine}. 
It enables ambient backscatter devices (AmBDs) to communicate by leveraging existing radio frequency (RF) signals, such as FM radio, cellular signals, Wi-Fi, TV broadcasts, Bluetooth, and LoRa, thereby obviating the need for dedicated RF emitters.
Specifically, AmBDs harvest energy from surrounding sources, such as RF signals, visible light, and mechanical vibrations, thus removing dependence on battery power and lowering maintenance costs~\cite{lopez2025zed}.  
By employing existing RF signals, AmBC benefits from the inherent sustainability and consistent availability of these ambient sources, thereby achieving energy-efficient and battery-free IoT operations.

Among various EH methods, visible light EH holds considerable promise owing to the widespread presence of lighting infrastructures indoors and abundant solar energy outdoors. 
Furthermore, modern lighting systems based on light-emitting diodes (LEDs) are capable of concurrently powering and communicating with IoT devices~\cite{haas2019survey}. This capability has attracted research interest, leading to the development of simultaneous lightwave information and power transfer~(SLIPT) technologies~\cite{ding2018slipt}. 
Within SLIPT systems, AmBDs can be simultaneously energized, accessed, woken up, and controlled via visible light communication (VLC). Compared with alternative EH approaches such as RF wireless power transfer (WPT) and mechanical vibrations, SLIPT-based VLC offers superior controllability, interference immunity, higher energy density, and simpler device management.

Capitalizing on the complementary advantages of VLC and AmBC, joint VLC-AmBC systems have emerged as a compelling paradigm for energy-neutral A-IoT networks~\cite{xie2024vlc}. Joint VLC-AmBC integration leverages the LED infrastructure for power delivery and data communication, alongside ambient RF sources (AmRFSs) for enhanced network coverage and robust device connectivity. 
Motivated by the substantial advantages of joint VLC-AmBC solutions, recent years have witnessed research efforts supporting technological and practical deployments~\cite{varshney2024lifibc,xie2024vlc,xie2024light,koskinen2025li2bc}. 
However, existing studies address different aspects separately, such as VLC/SLIPT design, AmBC mechanisms, or prototype demonstrations, rather than a unified system architecture. 
To bridge the gap between academic advancements and practical implementations, that is, the lack of a system-level and deployment-oriented perspective, this article provides a comprehensive perspective on joint VLC-AmBC technologies, highlighting the principles, architectures, practical implementations, and real-world applicability. 

The article first reviews the fundamentals of VLC, SLIPT, and AmBC, and then presents a joint VLC-AmBC system architecture with VLC access points (APs), energy-neutral AmBDs, AmRFS, and general-purpose RF receivers. 
The paper further discusses representative application scenarios, introduces three practical AmBD prototypes with proof-of-concept (PoC) experiments, and finally highlights deployment challenges together with promising future research directions.
\section{Overview of VLC, SLIPT, and AmBC}
\subsection{Visible Light Communication}
VLC has emerged as a promising wireless technology that utilizes the visible light spectrum, with wavelengths from approximately 380 to 750~nm, to achieve wireless communication. VLC systems modulate the intensity of emitted visible light from ubiquitous LED sources to convey data, offering significant benefits including additional bandwidth, immunity to RF interference, enhanced security, and license-free operation. These features make VLC particularly suitable for IoT applications, especially in dense indoor environments where conventional RF-based communication encounters limitations related to congestion, interference, and security vulnerabilities.

VLC enables connectivity and communication among IoT devices and infrastructure, thus facilitating applications such as visible light positioning (VLP), intelligent appliance control, and high-speed data transmission.
Moreover, VLC-based IoT devices can benefit from inherently secure communications owing to the confinement of optical signals within physical boundaries, significantly reducing risks associated with wireless eavesdropping and interference. 
VLC has been standardized in IEEE~802.11bb~(LiFi), 802.15.7, 802.15.13, and ITU-T~G.9991, showing a promising pathway to large-scale deployment and application in next-generation networks.
\vspace{-5pt}
\subsection{Simultaneous Lightwave Information and Power Transfer}
Extending the functionality of VLC, SLIPT technology integrates communication and EH capabilities within the same optical wireless link. In SLIPT systems, IoT devices simultaneously receive optical power for EH and decode modulated signals for data communication. Equipped with photovoltaic~(PV) cells or photodetectors, IoT devices can directly convert received optical signals into electrical energy, achieving battery-free, energy-neutral operations.
Compared with \mbox{RF-WPT} methods, such as far-field RF-EH, SLIPT provides higher energy conversion efficiency~\cite{ding2018slipt}, benefiting from the abundant and focused energy in visible light wavelengths. 
In addition, leveraging existing LED lighting infrastructure, SLIPT minimizes deployment complexity and reduces maintenance and operational costs. Consequently, the seamless integration of VLC and SLIPT presents an ideal solution for A-IoT environments, ensuring sustainable operation and ubiquitous connectivity of IoT devices.
\vspace{-10pt}
\subsection{Ambient Backscatter Communication}
BC is a low-power wireless communication technology where BDs transmit data by reflecting and modulating incident RF signals without synthesizing their own carriers. 
BDs can achieve microwatt-level power consumption because of their passive or semi-passive nature, making them suitable for A-IoT scenarios.
BC systems are typically classified into three configurations, namely \emph{monostatic}, \emph{bistatic}, and \emph{ambient}. 
In monostatic BC, a reader simultaneously emits carrier signals and receives backscattered signals. 
Bistatic BC employs separate RF transmitters and receivers, enhancing communication range and deployment flexibility. 
AmBC further advances BC by utilizing existing ambient RF signals, such as Wi-Fi and cellular signals as RF carriers, eliminating the need for a dedicated RF infrastructure and reducing IoT devices costs.

The development of BDs has progressed through sophisticated hardware designs, diverse modulation schemes, and improved EH capabilities~\cite{jiang2023survey}, demonstrating low hardware complexity, small form factors, cost-effectiveness, and high energy efficiency~\cite{song2022survey}. 
These advancements facilitate practical implementation and widespread adoption in diverse A-IoT use cases, including environmental monitoring, smart healthcare, intelligent transportation, and industrial automation.

In summary, both VLC and AmBC individually offer compelling benefits for energy-neutral IoT applications. However, their integration into the joint VLC-AmBC system creates synergies thanks to their complementary advantages, enabling sustainable, self-powered A-IoT networks.
\vspace{-10pt}
\section{Joint VLC-AmBC System Architecture}
\begin{figure*}
\vspace{-18pt}
\centering
\includegraphics[width=1.98\columnwidth]{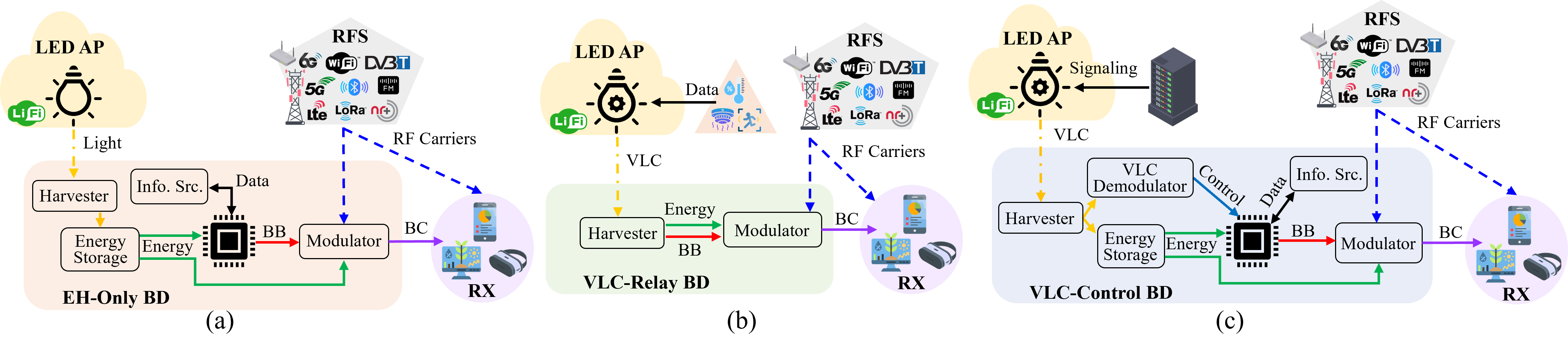}
\vspace{-12pt}
\caption{
Joint VLC-AmBC system architecture with three types of AmBDs:
(a) EH-Only AmBD harvests optical energy from LED APs solely for powering the device.
(b) VLC-Relay AmBD exploits VLC signals not only for EH, but also for relaying messages from VLC APs to RF receivers.
(c) VLC-Control AmBD uses VLC signals for EH and for controlling various operational states of the device, such as waking up, data collection, and backscattering.
All AmBDs modulate and backscatter ambient RF carriers from AmRFSs.
The backscattered signal can be received and decoded by general-purpose RF receivers.
}
\label{fig:ambd_scheme}
\vspace{-10pt}
\end{figure*}

As shown in Fig.~\ref{fig:ambd_scheme}, we propose the joint VLC-AmBC system architecture, integrating EH and BC. 
The architecture leverages existing LED lighting infrastructure as VLC APs for both illumination and data transmission, alongside AmBDs that harvest optical energy and utilize RF carriers to achieve energy-neutral operation. 
Furthermore, AmRFSs serve as providers of RF carrier signals, and general-purpose RF receivers act as endpoints for receiving and demodulating backscattered signals.
The following subsections detail each key component of the proposed system.
\vspace{-10pt}
\subsection{VLC Access Points}
VLC APs serve as light transmitters in the proposed system, playing a critical role in both providing illumination and transmitting data. 
VLC APs typically employ LED arrays as optical transmitters, leveraging their capability for fast intensity modulation to encode and transmit information. 
The principle of operation for VLC APs relies on intensity modulation and direct detection (IM/DD), where the LED transmitter modulates light intensity according to the data that needs to be transmitted from sensors and central controllers.
Common configurations incorporate dimming control and flicker mitigation for visual comfort, and spatial diversity strategies to enhance communication reliability and coverage. 
The deployment of VLC APs involves practical considerations, including precise placement for optimal illumination and communication coverage, management of optical interference, and periodic calibration and replacement of aging LED components.
\vspace{-10pt}
\subsection{Ambient RF Sources}
AmRFSs provide RF carrier signals required by AmBDs to enable communications. Available AmRFSs include widely deployed RF infrastructures, such as cellular base stations, Wi-Fi APs, FM radio stations, TV broadcast towers, Bluetooth transmitters, LoRa gateways, and DECT~NR+ emitters. 
By spanning diverse frequency bands and modulation standards, AmRFSs offer robust, stable, and ubiquitous carrier availability, significantly contributing to the effectiveness and scalability of AmBC within IoT ecosystems.
\vspace{-10pt}
\subsection{Ambient Backscatter Devices}
AmBDs constitute core components within the proposed system, enabling energy-neutral operation by harvesting energy from VLC and using ambient RF signals as communication carriers. Based on their functionalities and interactions with VLC, AmBDs can be categorized into three distinct types, which are described as follows and summarized in Table~\ref{tab:AmBD_keycompare}.
\begin{table*}[t]
\centering
\caption{Key feature comparison of typical AmBD types}
\vspace{-5pt}
\label{tab:AmBD_keycompare}
\renewcommand{\arraystretch}{1.2}
\setlength{\tabcolsep}{5pt}
\begin{tabularx}{\textwidth}{@{}>{\raggedright\arraybackslash}p{0.16\textwidth}
                            >{\raggedright\arraybackslash}p{0.20\textwidth}
                            >{\raggedright\arraybackslash}p{0.26\textwidth}
                            >{\raggedright\arraybackslash}p{0.30\textwidth}@{}}
\toprule
\textbf{Feature of AmBD} & \textbf{EH-Only} & \textbf{VLC-Relay} & \textbf{VLC-Control} \\
\midrule
Core role & Battery-free sensing & VLC-to-RF payload relay & Commanded sensing \\
Downlink from VLC & Not required & Required for relay & Required for control \\
Backscatter drive & MCU baseband modulation & AC component of VLC waveform & MCU baseband depending on VLC signaling \\
Relative complexity & Low & Low & Moderate \\
Key advantage & Simple and scalable & Extends VLC coverage to RF receivers & Fine control, duty cycling, locality \\
Main limitation & No downlink control & Relay only, no local sensing & Higher design effort; requires stable VLC links \\
Typical scenarios & Environmental sensing & Lighting/signage data relay & Healthcare rooms and secure areas \\
\bottomrule
\end{tabularx}
\vspace{-15pt}
\end{table*}
\subsubsection{\textbf{EH-Only AmBDs}}
This type of AmBD harvests optical energy solely for powering the device. Typically, EH-only BDs integrate energy harvesters and storage units, microcontroller units (MCUs), information sources, backscatter modulators, and RF antennas. 
The operational principle involves PV cells converting received optical signals into electrical energy, where a portion of this energy directly powers immediate device operations, and the remaining energy is stored in supercapacitors for utilization during low-light conditions.
The MCU collects and processes data from information sources, such as sensors, through analog-to-digital conversion, subsequently encoding the data into baseband~(BB) signals after analog-to-digital conversion. When ambient RF signals arrive at the device antenna, the BB signals modulate the impedance of the antenna, altering reflected signals accordingly. The backscattered signals thus carry information originating from the information source, facilitating the IoT data transmission.
\subsubsection{\textbf{VLC-Relay AmBDs}}
This type of AmBD exploits VLC signals not only for EH but also for relaying messages from VLC APs to RF receivers. Such AmBDs primarily include energy harvesters, backscatter modulators, and RF antennas. The energy harvesters convert incident VLC signals into electrical signals comprising both alternating current~(AC) and direct current~(DC) components. The DC component energizes device circuitry, while the AC component, with the embedded VLC data, is utilized to control backscatter modulation.
The AC signal modulates ambient RF carriers by adjusting the load of the RF antenna, thereby embedding VLC messages into backscattered signals, which can be received by conventional RF receivers. 
Such relay functionality extends the communication range and enhances VLC coverage in situations where the presence of obstacles prevents the direct reception of the VLC signal.
%
\subsubsection{\textbf{VLC-Control AmBDs}}
This type of AmBD utilizes VLC signals for EH and for actively controlling various operational states of the device, such as waking up, data collection, transmission, and sleeping. Typically, these devices comprise energy harvesters, VLC demodulators, MCUs, information sources, backscatter modulators, and RF antennas. 
The VLC receiver, in this case, processes the received optical wireless signals through filtering and amplification, subsequently decoding them within the MCU to extract embedded control instructions.
Upon receiving specific control signals, the MCU executes predetermined operations, such as initiating sensor measurements, processing and encoding sensor data, and performing baseband modulation. The resulting BB signals modulate ambient RF carriers via the backscatter modulator, enabling transmission of sensor data embedded within the backscattered signal. Such controllability enhances the operational flexibility of the devices.
\vspace{-10pt}
\subsection{General-Purpose RF Receivers}
General-purpose RF receivers are terminals that perform the reception and demodulation of backscattered signals from AmBDs.
The receivers typically incorporate RF front-end modules capable of capturing backscattered signals embedded into RF carrier waves. 
Moreover, they maintain versatility in receiving direct-path signals transmitted by AmRFSs, depending on specific communication schemes.
Practical implementations of RF receivers often require advanced signal processing capabilities to efficiently decode weak backscattered signals amid the presence of strong direct-path interference.
\vspace{-5pt} 
\section{Applications and Deployment Scenarios}
\begin{figure}
\vspace{-20pt}
\centering
\includegraphics[width=0.88\columnwidth]{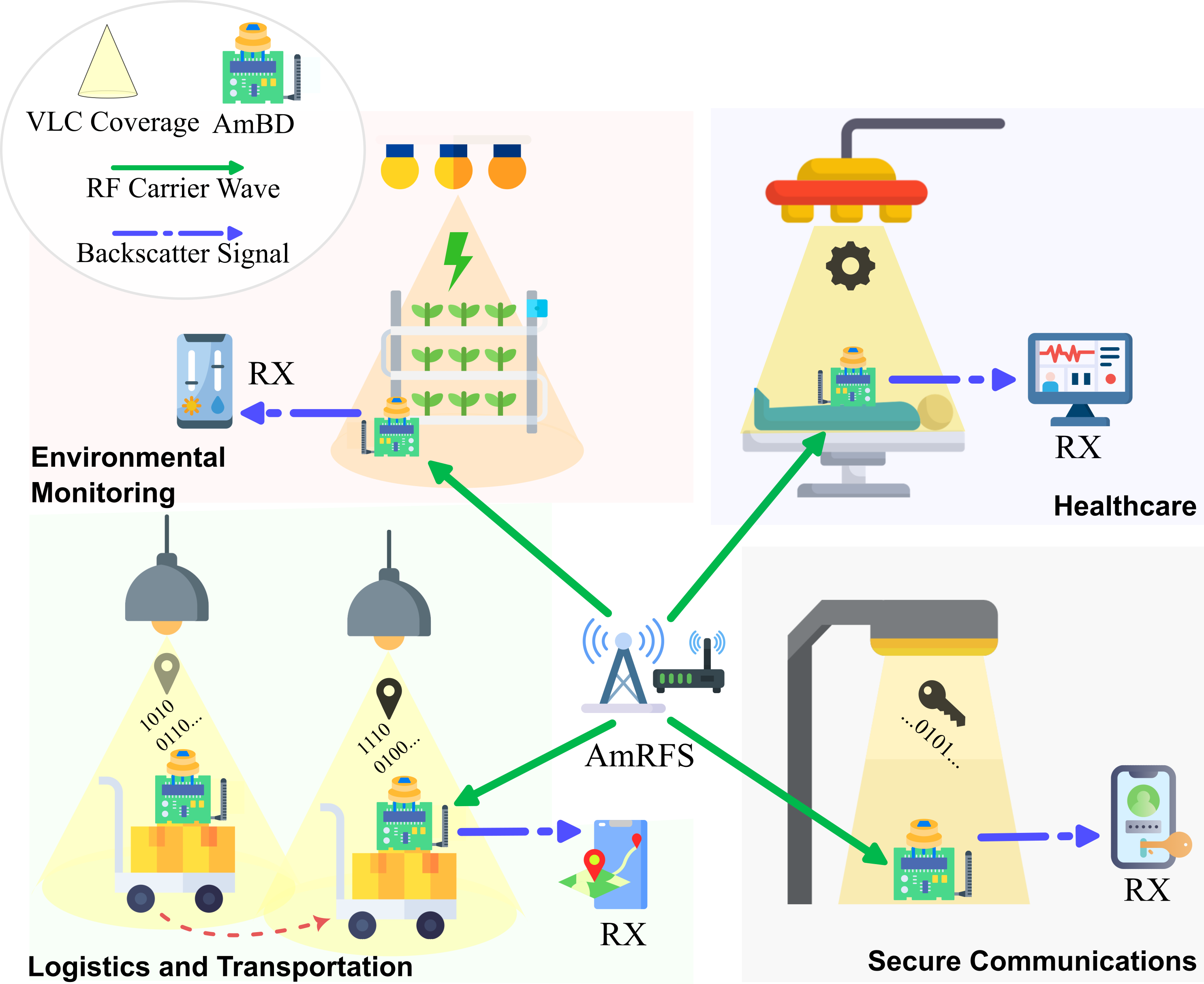}
\caption{Applications of the proposed system: Environmental monitoring, healthcare, logistics and transportation, and secure communications.}
\label{fig:application}
\vspace{-15pt} 
\end{figure}

In Fig.~\ref{fig:application}, we identify four typical deployment scenarios that can be supported by the proposed joint VLC-AmBC system, including environmental monitoring, healthcare, logistics and transportation, as well as secure communications.
\vspace{-10pt}
\subsection{Environmental Monitoring}
The joint VLC-AmBC systems can be used for environmental monitoring in diverse applications, including smart agriculture and industrial IoT. 
In smart agriculture scenarios, AmBDs can be deployed extensively in farmland areas, powered by abundant solar radiation during the daytime and artificial lighting at night. 
These AmBDs continuously monitor soil moisture, nutrient content, temperature, humidity, and crop health parameters without relying on battery replacements or dedicated power lines.
Similarly, for industrial IoT, AmBDs can be integrated into factory environments to monitor air quality, pollutant concentrations, machinery vibration, and temperature conditions. 
Leveraging existing lighting infrastructures for EH and data communication, these systems facilitate large-scale deployment, reduce installation complexity, and enable real-time monitoring, substantially enhancing operational efficiency and environmental compliance.
\vspace{-10pt}
\subsection{Healthcare}
In healthcare scenarios, the joint VLC-AmBC systems present opportunities for monitoring patient well-being and enhancing medical facility operations. 
For instance, in a hospital room, a wearable AmBD attached to a wristband can harvest energy from the bedside luminaire, receive VLC control signals for sensing activation, and backscatter physiological data such as heart rate and body temperature to an RF reader.
The room-confined VLC link enables localized AmBD wake-up and control with low interference to nearby patients and equipment, while AmBC reduces maintenance by avoiding frequent battery replacement and the use of a dedicated RF transmitter at the wearable device.
In addition, AmBDs embedded within hospital rooms, medical equipment, and medication containers can track asset usage, medication adherence, and environmental conditions such as room temperature, lighting intensity, and hygiene compliance. These capabilities help to improve patient care quality, reduce clinical errors, and streamline asset management, contributing to the management of healthcare facilities and the safety of patients.
\vspace{-10pt} 
\subsection{Logistics and Transportation}
The joint VLC-AmBC system also exhibit potential applications related to logistics, transportation, and asset tracking. 
Energy-neutral AmBDs can be seamlessly attached to packages, pallets, or containers, enabling comprehensive monitoring of asset locations, environmental conditions, and transportation routes. 
By harnessing available light energy during transit and within warehouses, these AmBDs can provide continuous, battery-free data transmission and status updates. 
Furthermore, the system can facilitate real-time inventory tracking and condition monitoring, such as detecting temperature-sensitive goods or vibration-sensitive equipment during transportation. 
These capabilities enhance transparency, reliability, and efficiency in supply chain management, mitigating losses and improving overall logistics performance.
\vspace{-10pt} 
\subsection{Secure Communications}
Leveraging the inherent spatial confinement of VLC signals, joint VLC-AmBC systems provide enhanced communication security through location-restricted communication areas. 
This feature is advantageous for applications requiring secure information transmission, such as secret key generation, user authentication, and secure data exchange in sensitive areas~\cite{zhang2025authscatter}. 
The VLC-assisted design can help restrict communication to specific, physically controlled locations， reducing vulnerability to unauthorized access and eavesdropping at the physical layer. Furthermore, utilizing VLC channels for secure initial key distribution, as well as employing ambient RF backscatter for subsequent data communication, provides a hybrid secure communication framework. 
This dual-mode system further enhances authentication robustness and security, supporting critical applications in government facilities, financial institutions, and private enterprise environments.
\vspace{-5pt} 
\section{Case Study: Joint VLC-AmBC System with AmBDs}
\begin{figure*}
\vspace{-15pt}
\centering
\includegraphics[width=1.98\columnwidth]{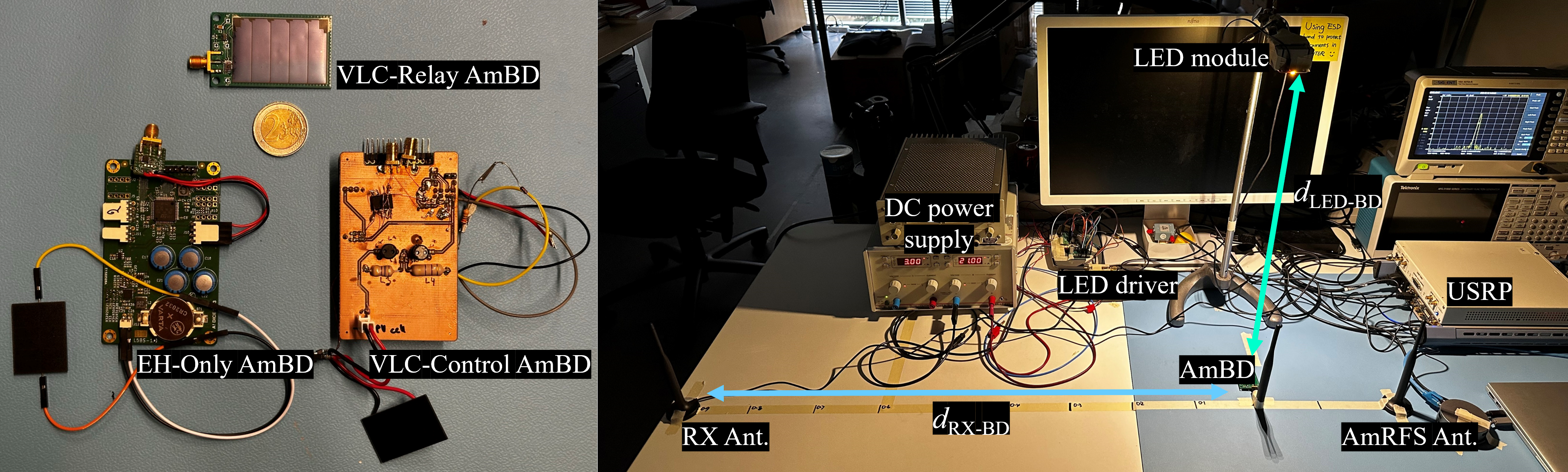}
\caption{Three PoC prototypes (left) for EH-Only, VLC-Relay and VLC-Control AmBDs, respectively, where a coin is placed to showcase their scales.
In the setup (right), VLC link distances $d_{\textrm{LED-BD}}$ and BC link distances $d_{\textrm{RX-BD}}$ are systematically varied during experiments using the controlled-variable method.
}
\label{fig:setup}
\vspace{-20pt} 
\end{figure*}

We present PoC implementations of the three types of AmBDs, as shown in Fig.~\ref{fig:setup}. Moreover, a series of end-to-end communication experiments is conducted to verify the implementation feasibility of the joint VLC-AmBC system.
\vspace{-20pt}
\begin{figure*}
  \centering
  \subfloat[]{\includegraphics[width=0.49\columnwidth]{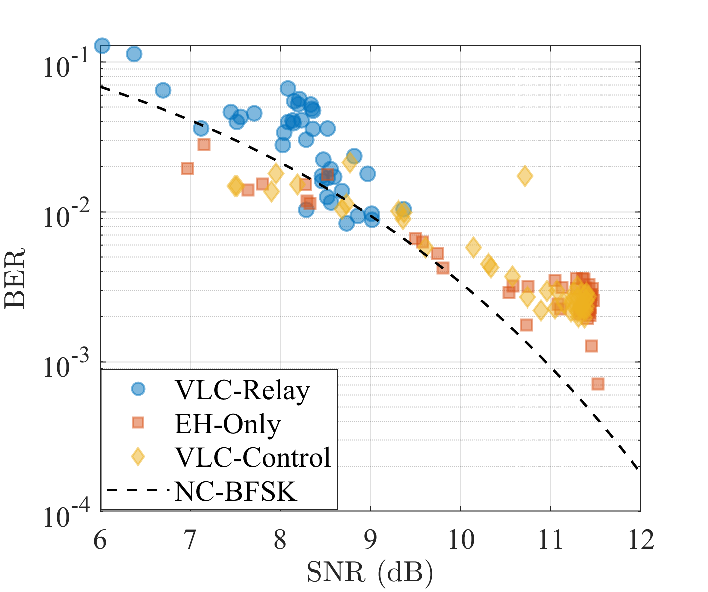}
   \label{fig:snr_ber_all_bd}}
    \subfloat[]{\includegraphics[width=0.49\columnwidth]{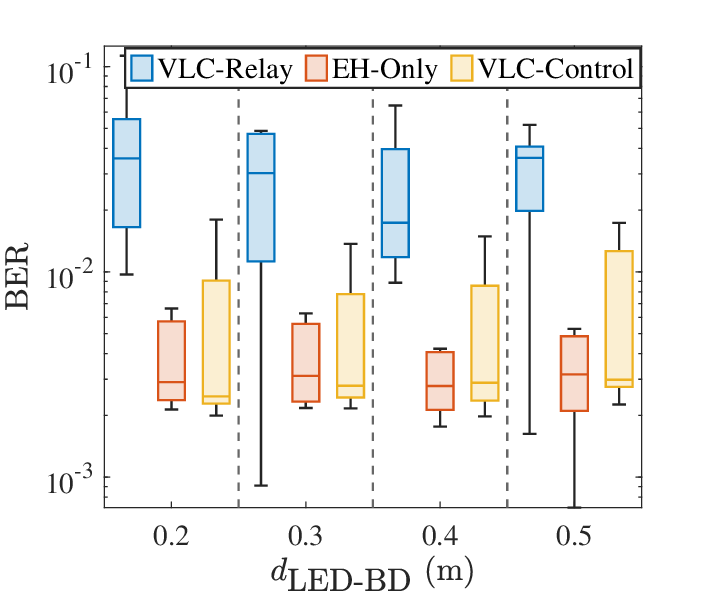}
   \label{fig:avg_ber_dledbd}}
  \subfloat[]{\includegraphics[width=0.49\columnwidth]{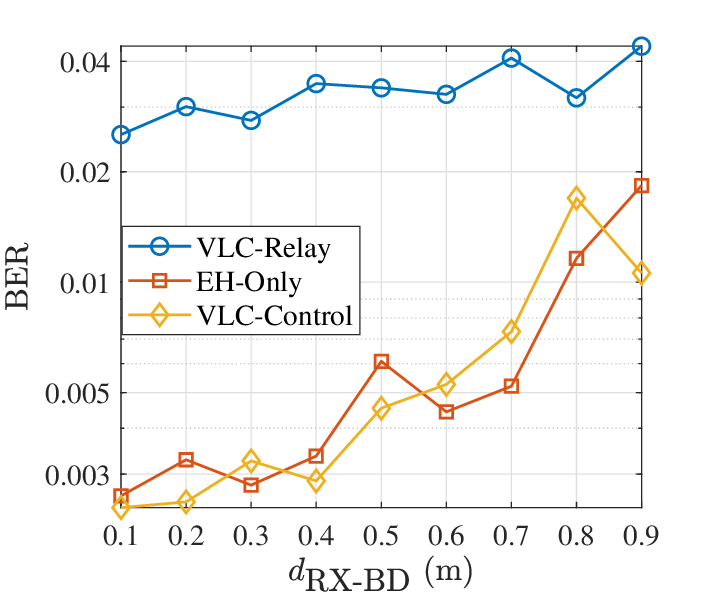}
   \label{fig:avg_ber_drxbd}}
  \subfloat[]{\includegraphics[width=0.49\columnwidth]{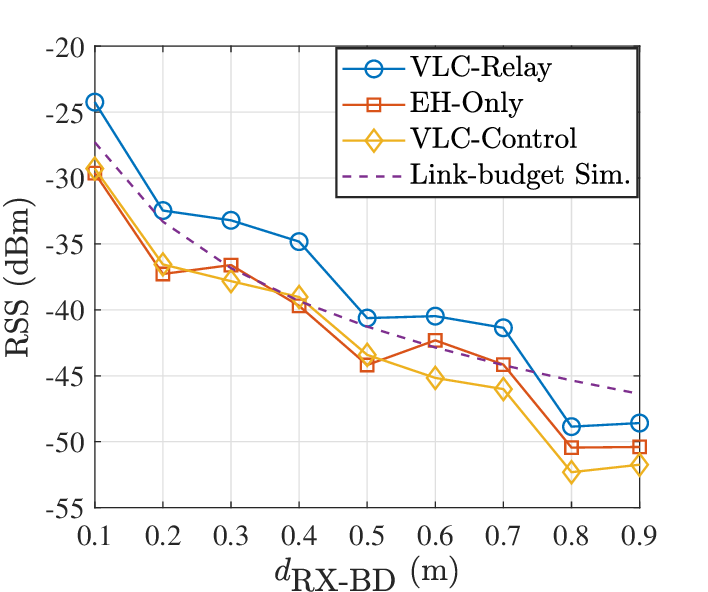}
   \label{fig:avg_rss_drxbd}}
  \vspace{-5pt} 
  \caption{
  Experimental results for the AmBDs. 
  (a) BER vs. SNR, compared with theoretical performance of a non-coherent BFSK system. 
  (b) BER across varying VLC link distances.
  (c) BER across varying BC link distances.
  (d) RSS across varying BC link distances, compared with the theoretical link budget~\cite{griffin2009linkbudget}.
  }
  \label{fig:exp_result}
\vspace{-15pt} 
\end{figure*}
\subsection{Experimental Setup and Prototypes}
\textbf{LED AP:} The VLC AP employs an array comprising seven phosphor-converted white LEDs OSRAM~GW~J9LHS2.4M. 
A Raspberry Pi RP2040 microcontroller generates light-modulating signals to control the LED array. Specifically, the signals adopt Manchester encoding coupled with binary frequency-shift keying (BFSK) modulation. 
Frequencies of BFSK modulation are set to 6~kHz and 8~kHz, respectively, deliberately chosen above 2~kHz to prevent visible flicker from affecting illumination quality. 
The BFSK symbol duration is 0.5~ms, corresponding to a symbol rate of 2000~symbols/s. With Manchester encoding, the data rate is 1~kbps.
The transmitted frame structure includes a preamble, consisting of a 7-bit Barker sequence `1110010' for signal detection and synchronization, followed by an 18-bit pseudorandom payload sequence. 
Furthermore, the LED driving stage incorporates an LED driver module using the chip OnSemi CAT4109, which biases the modulated signals with DC current from a constant power supply.

\textbf{AmRFS:} The RF source consists of a Rohde \& Schwarz SGT100A signal generator configured to emit a continuous sine-wave carrier at a frequency of $f_{\rm c}=2.4$~GHz. During experiments, the transmission power was varied from -25~dBm to 25~dBm in increments of 5~dBm, facilitating comprehensive performance analysis under diverse ambient RF conditions. This controlled source is adopted to provide a repeatable baseline for end-to-end validation. In practical deployments with ambient Wi-Fi or cellular signals, the received power of the illuminating RF carrier might be time-varying and intermittent due to traffic activity, channel access, and the mobility of AmBDs. Therefore, the experiments should be interpreted as implementation feasibility validation, while robust practical operation would require energy buffering and adaptive duty cycling at AmBDs, carrier-aware backscatter operation, and receiver-side interference suppression.

\textbf{AmBDs:} Three AmBD prototypes are developed and evaluated in distinct operational configurations, shown in Fig.~\ref{fig:setup}.

The {\em EH-Only AmBD prototype} integrates a solar panel ANYSOLAR SM500K12L, an energy harvester Analog Devices ADP5091 with four 0.1~F supercapacitors, an MCU STMicroelectronics STM32L562, a backscatter modulator based on an RF switch Analog Devices ADG919, and an RF antenna. 
The solar panel harvests optical energy, subsequently converted by the energy harvester into electrical power, partially utilized immediately for device operation and partially stored within supercapacitors for usage during insufficient lighting conditions. 
The MCU employs a low-power timer to generate pulse-width modulation (PWM) signals for baseband modulation, implementing BFSK by alternating the PWM frequency between 6~kHz and 8~kHz while maintaining a fixed duty cycle of 50\%. The backscatter modulator, implemented via the RF switch connecting the antenna, alternates the antenna termination between \emph{open} and \emph{short} based on the MCU-generated baseband signals. Thus, the ambient RF carrier illuminating the antenna is modulated accordingly and backscattered to the RF receiver.

The {\em VLC-Relay AmBD prototype} incorporates a solar panel TDK BCS4430B6, an energy harvester consisting of a low-pass filter (LPF) and comparator Texas Instruments TLV7031, a backscatter modulator based on the RF switch ADG919, and an antenna. 
The solar panel captures the VLC signals and converts them into electrical signals, which are further split into DC and AC components.
The DC component extracted through the LPF supplies power for device operation, while the comparator-regulated AC component, as the BB signal, directly controls backscatter modulation, carrying VLC-embedded information. 
By altering antenna reflection coefficients through load modulation, the incident RF carrier is modulated and backscattered.

The {\em VLC-Control AmBD prototype} consists of a solar panel ANYSOLAR SM710K12L, an energy harvester ADP5091, a transimpedance amplifier (TIA) based on an operational amplifier Texas Instruments TLC2262, an MCU STM32L562, a temperature sensor, a backscatter modulator based on the RF switch ADG919, and an RF antenna. 
The solar panel captures and converts VLC signals into electrical signals, subsequently amplified and decoded by the TIA and MCU, respectively. 
The decoded VLC signals provide control commands for further operations of the device, such as sensor activation and data acquisition. 
Upon receiving specific commands, the MCU obtains and processes sensor data, performs baseband modulation, and controls the backscatter modulation. 
Thus, the sensor data are embedded into the backscattered signal by modulating incident RF carriers.

We estimate average power budgets of the prototypes from implemented circuits and steady operating modes. The EH-Only, VLC-Relay, and VLC-Control prototypes consume approximately 30~$\mu$W, 1.08~$\mu$W, and 3.03~mW, respectively. 
The first two prototypes operate in the microwatt regime, whereas the VLC-Control prototype requires a higher budget mainly due to the VLC receiver front end with the TIA. Under direct illumination from the VLC AP in the PoC setup, all three prototypes operated without external power supplies.

\textbf{RX:} The receiver platform consists of a software-defined radio USRP X300, which is configured to capture RF signals at a sampling rate of $f_{\rm s}=200$~kHz. 
The captured RF samples are transferred to a host computer, where they undergo non-coherent FSK demodulation. 
Key metrics, including signal-to-noise ratio~(SNR), bit-error rate~(BER), and received signal strength (RSS), are recorded and analyzed to comprehensively evaluate the system performance.
\vspace{-5pt}
\subsection{Experimental Configuration}
To thoroughly assess the three proposed AmBD types within the joint VLC-AmBC architecture, performance evaluations are conducted by investigating the influence of both the VLC and BC links. 
As shown in Fig.~\ref{fig:setup}, two distances are systematically varied during experiments using the control variates method: the distance from the LED AP to the AmBDs, denoted as $d_{\textrm{LED-BD}}$, and the distance from the AmBDs to the RX, denoted as $d_{\textrm{RX-BD}}$. The distance $d_{\textrm{LED-BD}}$ is changed from 0.2 to 0.5~m, in steps of 0.1~m, while the distance $d_{\textrm{RX-BD}}$ is varied from 0.1 to 0.9~m, in steps of 0.1~m. This systematic variation facilitates understanding of how changes in positioning and proximity affect the overall performance of the joint VLC-AmBC links.
\vspace{-5pt}
\subsection{Experimental Results}
Fig.~\subref*{fig:snr_ber_all_bd} presents the communication performance of the three AmBDs, showing the BER versus SNR of the received backscatter signal. 
BER decreases significantly as SNR increases, aligning closely with the theoretical performance of non-coherent BFSK modulation, denoted by the dashed line. This result confirms the practical viability of the proposed system under realistic operational conditions.
Fig.~\subref*{fig:avg_ber_dledbd} investigates the effect of the VLC link on communication performance by plotting BER across varying the LED-BD distances. BER values remain relatively stable, indicating successful VLC reception within the distance ranges under evaluation. This stability highlights the feasibility of employing VLC for energy delivery and communication in the proposed system.
Fig.~\subref*{fig:avg_ber_drxbd} illustrates BER variations concerning the RX-BD distance. An increasing trend in BER with larger $d_{\textrm{RX-BD}}$ highlights the impact of RF signal attenuation on BC performance. The results show practical operational ranges for communication.
Fig.~\subref*{fig:avg_rss_drxbd} shows RSS performance over varying RX-BD distances, demonstrating a clear RSS decline with increased distances. The measured RSS closely matches theoretical predictions from the backscatter link-budget simulation~\cite{griffin2009linkbudget}, validating experimental accuracy and reinforcing the practicality of joint VLC-AmBC systems for integrated sensing and communication applications.
\begin{figure}
\vspace{-15pt}
  \centering
  \subfloat[]
{\includegraphics[width=0.48\columnwidth]{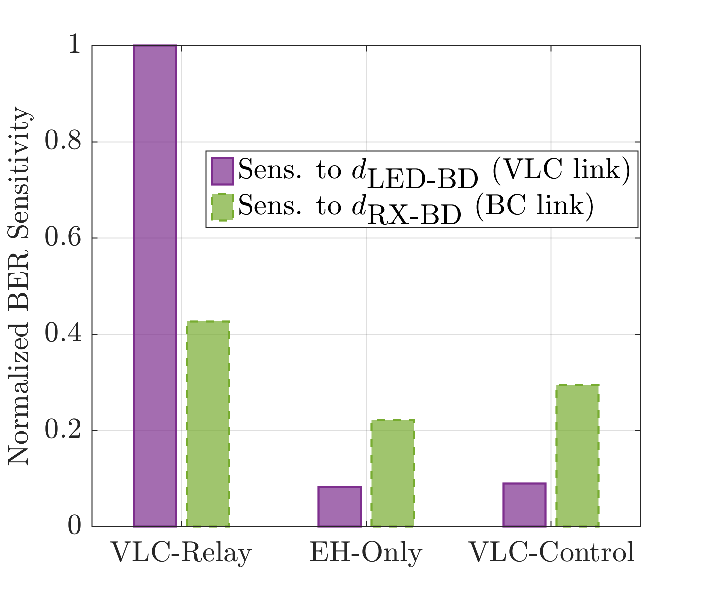}
   \label{fig:ber_sen}}
\subfloat[]
{\includegraphics[width=0.48\columnwidth]{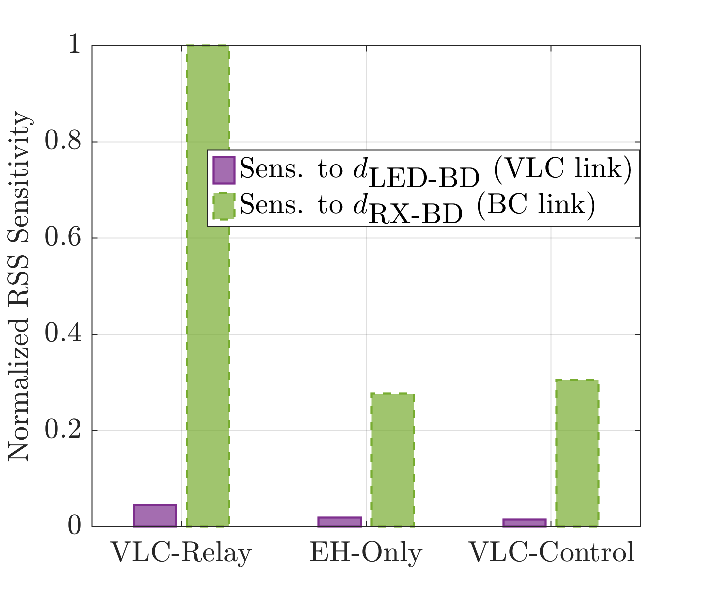}
   \label{fig:rss_sen}}
  \vspace{-5pt}
  \caption{
  Normalized sensitivity of (a) BER and (b) RSS, in terms of the distance changes in the VLC link and BC link per centimeter.
  }
  \label{fig:sensitivity}
\vspace{-15pt}
\end{figure}

Fig.~\subref*{fig:ber_sen} and~\subref*{fig:rss_sen} summarize the sensitivity of BER and RSS, respectively, in terms of the distance changes in the VLC link ($d_{\textrm{LED-BD}}$) and BC link ($d_{\textrm{RX-BD}}$) per centimeter. 
Changes in the VLC link dominate the BER performance for the VLC-Relay AmBD, whereas the BC link dominates the performance for EH-Only and VLC-Control AmBDs, due to the different operation mechanisms.
For the VLC-Relay AmBD, the VLC signals are PV-converted and used to modulate the RF carrier signal for backscatter, such that the backscattered signal inherits errors from the VLC link. 
However, for the other two AmBDs, the backscattered signal does not inherit the VLC errors.
Furthermore, the RSS is governed by the BC link for all three AmBDs.
This decoupling indicates that BER performance can be improved by optimizing the optical path, e.g., LED alignment and optical power, while ensuring adequate RSS requires focusing on backscatter geometry and RF power budgets. By tailoring design efforts to the primary limiting link for each metric, one can achieve more efficient trade-offs in the proposed system.
\vspace{-5pt}
\subsection{Key Insights from Prototypes and Experiments}
The three prototypes reveal a trade-off between functionality and implementation complexity. 
The EH-Only AmBD provides the simplest design for battery-free sensing, yet it does not support VLC-based device control. 
The VLC-Relay AmBD achieves low hardware complexity for message forwarding, whereas its BER is more sensitive to the optical link since the backscattered signal directly inherits the quality of the received VLC waveform. 
The VLC-Control AmBD offers the highest flexibility for commanded sensing and duty cycling, but this benefit comes with higher circuit complexity and a tighter power budget due to VLC signal amplification, decoding, and MCU processing. 
In addition, the experiments indicate that RSS is mainly governed by the BC link, whereas BER sensitivity depends on the specific operating mechanism of each prototype. 
Implementations also shows that stable optical alignment, sufficient illumination, and sufficient ambient RF signal strength are important to maintain device operation.
\vspace{-5pt}
\section{Deployment Challenges and Future Directions}
The following discussion summarizes major deployment challenges and highlights promising research directions.
\subsection{Deployment Challenges in Practical VLC-AmBC Systems}
Practical deployment of joint VLC-AmBC systems faces several challenges. First, VLC links are sensitive to line-of-sight blockage, device orientation, and illumination non-uniformity, which can reduce harvested energy and communication reliability. 
Second, coexistence with lighting infrastructure requires communication functions to remain compatible with illumination constraints, such as dimming control and luminaire placement. 
Third, ambient RF carriers used by AmBDs are inherently variable across time, frequencies, and positions in the service area, while RF receivers need to handle direct-path interference and coexistence with other wireless systems. 
In extreme conditions with insufficient light or weak ambient RF signals, continuous operation may require extra resources.
In such cases, practical robustness relies on graceful degradation through energy storage on supercapacitors, adaptive duty cycling, source or channel selection when multiple ambient carriers are available, and low-complexity fallback operation with reduced sensing or reporting frequency.
\subsection{Integrated VLC-AmBC System with Sensing}
The joint VLC-AmBC system holds potential to evolve toward integrated sensing and communication (ISAC) frameworks. 
By leveraging inherent characteristics of visible light and ambient RF signals, VLC-AmBC systems can facilitate simultaneous high-resolution indoor positioning, environmental sensing, and data transmission. 
Future research can focus on integrating advanced sensing techniques such as RF fingerprinting, channel charting via visible light positioning, and multipurpose sensing using ambient RF signals. 
Developing robust algorithms for simultaneous sensing and communication and addressing challenges, such as interference management and synchronization, will enhance the versatility of VLC-AmBC systems for smart and connected environments.
\subsection{Optimization of VLC Deployment and Energy Efficiency}
Optimizing the deployment of VLC infrastructures presents an important avenue for improving the performance and energy efficiency of the proposed system. Future studies can investigate adaptive lighting design strategies, optimized LED placement, and intelligent dimming control methods to maximize both EH and communication performance. 
Moreover, the integration of machine learning algorithms and data-driven techniques can further enhance adaptive energy management and resource allocation in VLC systems. Addressing practical deployment constraints, such as varying illumination needs, user mobility, and coverage optimization, will also be essential for system implementation in real-world A-IoT scenarios.
\subsection{Large-scale Deployment of AmBDs in A-IoT Networks}
Large-scale deployment of numerous AmBDs within dense and confined areas introduces critical challenges regarding resource management, multiple access protocols, interference mitigation, and efficient device addressing. Given the limited available bandwidth and power resources, novel approaches for adaptive and dynamic resource allocation and device coordination are required. 
Research can focus on scalable and energy-efficient communication protocols, innovative multiple access schemes, and sophisticated addressing methods to effectively manage extensive AmBD networks. 
Addressing these challenges is essential for achieving seamless interoperability, maintaining high communication quality, and ensuring energy-neutral operation in large-scale A-IoT deployments.
\section{Conclusion}
In this article, we have proposed the joint VLC-AmBC system architecture, demonstrating its potential to support energy-neutral A-IoT implementations.
By integrating VLC with AmBC, the system enables AmBDs to operate sustainably, harvesting ambient energy and communicating effectively without the need for dedicated power sources. 
We have provided detailed insights into the operational principles and implementation of the three AmBD types.
Experimental results confirmed the feasibility of AmBDs under realistic scenarios. 
Furthermore, we have identified promising future research directions and discussed critical open issues. 
Addressing these challenges will advance the joint VLC-AmBC technology, thereby fostering its widespread adoption and paving the way to sustainable, efficient, and scalable A-IoT solutions.

\bibliographystyle{IEEEtran}
\bibliography{reference}

\clearpage

\vfill

\end{document}